\documentclass[prb,twocolumn,showpacs,preprintnumbers,amsmath,amssymb]{revtex4}

\usepackage{color}
\usepackage[utf8]{inputenc}
\usepackage[dvips]{graphicx}
\usepackage{wrapfig}
\usepackage{amsmath}
\usepackage{subfigure}
\usepackage{slashbox}

\newcommand{\bra}[1]{\langle #1|}
\newcommand{\ket}[1]{|#1\rangle}

\begin{document}
	
	\title{Quasiparticles in the Kondo lattice model at partial fillings of the conduction band}
	
	\author{Sebastian Smerat}
		\email{smerat@physik.rwth-aachen.de}
		\affiliation{Institut für theoretische Physik C, RWTH Aachen University, D-52056 Aachen, Germany}
                \affiliation{JARA-Fundamentals of Future Information Technology}
	\author{Ian P. McCulloch}
		\affiliation{School of Physical Sciences, University of Queensland, QLD 4072, Australia}
	\author{Herbert Schoeller}
		\affiliation{Institut für theoretische Physik A, RWTH Aachen University, D-52056 Aachen, Germany}
                \affiliation{JARA-Fundamentals of Future Information Technology}
	\author{Ulrich Schollwöck}
		\affiliation{Institut für theoretische Physik C, RWTH Aachen University, D-52056 Aachen, Germany}
                \affiliation{JARA-Fundamentals of Future Information Technology}
	\date{\today}
	\begin{abstract}
		We study the spectral properties of the one-dimensional Kondo lattice model as function of the exchange coupling, the band filling, and the quasimomentum in the ferromagnetic and paramagnetic phase. Using the density-matrix renormalization group method, we compute the dispersion relation of the quasiparticles, their lifetimes, and the $Z$-factor. As a main result, we provide evidence for the existence of the spinpolaron at partial band fillings. We find that the quasiparticle lifetime differs by orders of magnitude between the ferromagnetic and paramagnetic phase and depends strongly on the quasimomentum.
	\end{abstract}
	\maketitle

	\section{Introduction}
	The Kondo lattice model (KLM) has been a matter of constant interest for more than the last three decades. In two and three dimensions it is one of the common models to describe heavy-fermion \cite{Coleman2007} physics and is also a possible candidate for high-T$_c$ superconductivity.\cite{Tsunetsugu1997} Our motivation to study the one-dimensional KLM \cite{Tsunetsugu1997} is threefold. Firstly it has been shown \cite{Reininghaus2006} that the spin polaron, which is a quasiparticle of the KLM, plays an important role in nonequilibrium transport in a quantum wire coupled to a ferromagnetic spin chain; our method provides the possibility to investigate the quasiparticles of the model. The spin polaron might also play an important role in the electron spin decay process \cite{Schliemann2003} in quantum dots induced by the hyperfine interaction due to nuclear spins. Secondly it might be helpful to understand the one-dimensional model in greater detail to assist investigations in higher dimensions. And lastly the model has become interesting for the description of mesoscopic systems, like carbon nanotubes filled with fullerenes or endohedral fullerenes, so called peapods.\cite{Krive2006} The aim of this work is to expand on the understanding of the spectral properties of the 1d KLM. We show, by means of the density-matrix renormalization group \cite{White1992,Schollwoeck2005}  (DMRG), that persistent quasiparticle states exist, namely the spin polaron states, and extrapolate their lifetimesand their weights. Furthermore we calculate the quasiparticle dispersion relation. For the case of half-filling we show that our results qualitatively agree with the results of a strong coupling expansion in Ref. \onlinecite{Trebst2006}. We compare dispersion relations and confirm the existence of a critical coupling constant at which the effective quasiparticle mass diverges for large momenta.
	
\begin{figure}[b]
		\begin{center}
			\includegraphics[width=8.6cm]{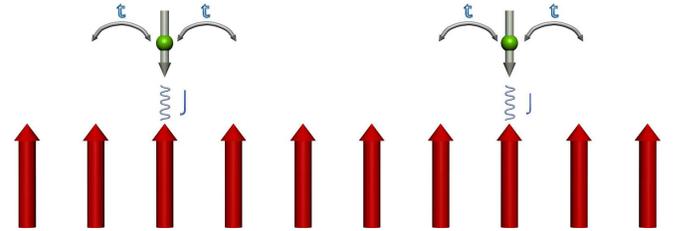}	
			\caption{\emph{(Color online)} The Kondo lattice model. The conduction electrons are depicted in the upper row (green) and the localized electrons are depicted as bolt arrows in the lower row (red).}
			\label{fig::KLM}
		\end{center}
	\end{figure}

	The KLM (Fig. \ref{fig::KLM}) consists of a lattice with one localized f-electron on every of the $L$ lattice sites, which do not interact with each other and a band of itinerant conduction electrons of finite filling $n$, coupled to the localized electrons by an on-site Heisenberg exchange interaction. The Hamilton operator of the 1d KLM is given by
	\begin{eqnarray}
	 		H = -t \sum_{i=1}^L \sum_{\sigma=\uparrow,\downarrow} \left( c_{i \sigma}^{\dagger} c_{i+1 \sigma}+\mbox{h.c.} \right) + J \sum_{i=1}^L \mathbf{S}_i \cdot \mathbf{s}_i,
	\end{eqnarray}
	where $t$ is the hopping parameter, $c_{i \sigma}^{\dagger}$ generates an electron at site $i$ with spin $\sigma$ and $\mathbf{S}_i$ ($\mathbf{s}_i$) are the spin operators of the
        localized (conduction) electrons at site $i$, respectively. $J$ is the Kondo coupling constant; we will consider
        only $J>0$ here, i.e. the antiferromagnetic coupling case. With $k$ we denote the quasimomentum in the following.
        
	In principle, the 1d KLM supports three phases, depending on the filling $n$ and on the coupling $J$: A ferromagnetic, a paramagnetic and, at half-filling ($n=1$) only, a spin liquid phase, see Fig. \ref{fig::phase_diagram}. At half-filling of the conduction electron band the model is best understood and early works using large-$N$-methods \cite{PhysRevB.30.3841} and the Gutzwiller approximation \cite{PhysRevB.34.6420,PhysRevLett.55.995} revealed that the magnetic properties are due to the competition of the RKKY interaction and the formation of Kondo singlets, where such a singlet is a conduction electron forming a spin singlet with a localized electron. Due to half-filling, the electrons induce an effective RKKY interaction between the localized spins, which forces antiferromagnetic power-law correlations in the ground state. The occurence of RKKY oscillations or $2k_F$-oscillations could be confirmed in Ref. \onlinecite{Yu1993} using DMRG. By means of exact diagonalization \cite{Tsunetsugu1992} and quantum Monte Carlo \cite{PhysRevLett.65.3177} it was shown that the ground state is spin- and charge-gapped and that it is a singlet of total spin. Therefore the ground state can be associated for all $J$ with the universality class of spin liquids. There has been a controversial discussion about the size of the Fermi volume (which is a single line in one dimension), whether it is small and therefore the Fermi wave vector is $k_{F} = \frac{\pi}{2}n$ or whether it is large and therefore $k_{F} = \frac{\pi}{2} \left( n + 1 \right)$. While a small Fermi volume would correspond to only conduction electrons contributing to the Fermi volume, a large Fermi volume would mean that the localized electrons also contribute to the Fermi volume. The idea of a large Fermi volume is borrowed from the periodic Anderson model.\cite{Tsunetsugu1997} There the f-electrons can move back to the conduction band and therefore contribute to the Fermi volume. The KLM can be derived from the periodic Anderson model \cite{Schrieffer1966} in the case of large Coulomb interaction, where only one localized electron per site is allowed and other occupations are fully supressed. This gives rise to the question whether the Fermi volume is also large in the KLM. Lately the authors of Ref. \onlinecite{Trebst2006} could argue within a strong coupling expansion and from the evaluation of the conduction electron density that the Fermi volume in the case of half-filling is small. In the same work, Ref. \onlinecite{Trebst2006}, the quasiparticle dynamics of the half filled KLM have been examined as well. It has been possible to calculate the quasiparticle dispersion relation to good accuracy, where the quasiparticle mass has been found to diverge around $k \approx \pi$ for $J > J_c \approx 0.50 \pm 0.02$. Therefore the quasiparticles behave like nearly localized f-electrons due to the strong correlation of the conduction and localized electrons.
	
	\begin{figure}
		\begin {center}
			\includegraphics[width=8.6cm]{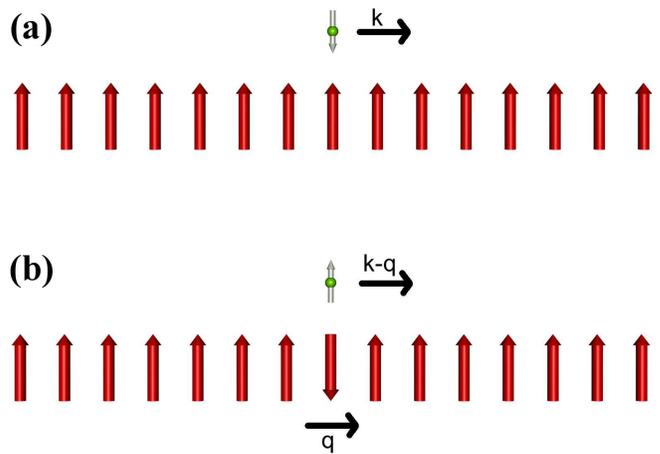}
		\end{center}
		\caption{\emph{(Color online)} Representatives of the constituent elements of the spin polaron: a) Electron spin down; b) Electron spin up and one of the lattice spins down.}
		\label{fig::spinpolaron}
	\end {figure}
	In the limiting case of vanishing conduction electron density it could be rigorously shown \cite{Sigrist1991} by both applying the Perron-Frobenius theorem and later exact diagonalization \cite{Sigrist1992a} that the KLM is ferromagnetic for all $J$.
	Importantly, Sigrist \cite{Sigrist1991} could show, that the quasiparticle of the Kondo lattice model is the spin polaron, which corresponds to an excited state separated from a continuum of scattering states. Representatives of the constituent elements of the spin polaron are shown in Fig. \ref{fig::spinpolaron}. In Fig. \ref{fig::spinpolaron}(a) the localized spin lattice is completely ferromagnetic and the electron spin is oriented in the opposite direction. Due to the antiferromagnetic exchange interaction the electron energy for quasimomentum $k$ is reduced. Via spin-flip processes, this state is coupled to the states shown in Fig. \ref{fig::spinpolaron}(b), where the electron spin and one of the localized spins are flipped and the momentum $q$ has been transferred to the spin lattice. These are states of higher energy since the antiferromagnetic interaction can only reduce the energy if the two flipped spins are at the same site. The coupling leads to a level repulsion between the states of Fig.\ref{fig::spinpolaron}(a) and Fig.\ref{fig::spinpolaron}(b), the energetically lower one corresponding to the spin polaron state and the higher ones forming the scattering states band. As proposed in Ref. \onlinecite{Reininghaus2006}, the spin polaron state is expected to have a very long life-time if its energy lies outside the band of scattering states, so that it is protected against magnon absorption and emission processes.

	At partial band fillings $n$, ferromagnetism also survives in the strong-coupling limit,\cite{Sigrist1992} where the KLM can be mapped to an effective Heisenberg model with a ferromagnetic exchange coupling. In this limit the formation of Kondo singlets, which move through the lattice, is sufficient to explain the occurence of ferromagnetism, but this does not exclude RKKY interaction, which might still play an important role. From exact diagonalization studies \cite{PhysRevB.47.8345} it follows that the KLM is ferromagnetic from a small but finite $J(n)$ up to $J \rightarrow \infty$ for all $n$ except half-filling. This raises the question, which mechanism drives ferromagnetism at intermediate couplings $J$ and one proposal \cite{PhysRevLett.78.2180} is that double exchange might be the crucial mechanism, where one conduction electron is responsible for screening several localized electrons. Screening is energetically effective in the antiferromagnetic KLM as long as $J(n)$ surmounts a critical value and forces the localized spins to align in the same direction.
	
	At a certain $J(n)$ the KLM approaches a second order transition \cite{PhysRevB.65.052410} by lowering $J$ to a paramagnetic phase, where the spin polaron \cite{Methfessel1968} might play an important role. The transition line has been calculated using exact diagonalization \cite{PhysRevB.47.8345} and has been refined later by means of bosonization.\cite{PhysRevLett.78.2180} The destruction of the ferromagnetic phase is described by a quantum random transverse-field Ising Hamiltonian.\cite{PhysRevLett.78.2180} Approaching the transition line from high $J$ it has been proposed \cite{PhysRevLett.78.2180, McCulloch2001} that the large ferromagnetic cluster splits up in several small clusters each corresponding to one spin polaron. Just below the transition line the small clusters' direction of magnetization is not the same anymore for all clusters and leads to zero net magnetization. By means of DMRG the spin structure factor of the localized electrons could be calculated \cite{McCulloch2001} and it has been found that the size of the Fermi volume is small for very low $J$ and becomes large approaching the transition line from lower $J$. From this one can conclude that near the transition line, the localized electrons are incorporated in the Fermi volume and therefore spin polarons are formed. Lowering $J$ the spin polarons are destroyed.
	The paramagnetic KLM has also been argued \cite{Shibata1999} to belong to the class of Tomonaga-Luttinger liquids.\cite{Haldane1981} This is motivated by the gapless spin and charge excitation,\cite{Tsunetsugu1997} which also makes the model difficult to handle with numerical methods using finite system sizes in this regime. From an analysis of Friedel oscillations, which are $2 k_F$ or $4 k_F$ oscillations, the Luttinger parameters could be determined for $J>1.8$ and the Fermi volume has been found to be large.
	For very low $J$, RKKY or $2 k_F$ oscillations dominate the correlation functions of the KLM. This could be attributed \cite{PhysRevLett.78.2180} to the backscattering of the conduction electrons at the localized electrons.
		
	In this paper we consider the spectral properties of the Kondo lattice model at partial band fillings. We will calculate the dispersion relation in the ferromagnetic phase for different Kondo couplings $J$ and various fillings $n$ and show that the spin polaron state exists. We are able to confirm the results of Ref. \onlinecite{Trebst2006} at half-filling. In a second step we will show from the width of the spectral densities that the lifetime of the spin polaron is very long and therefore the quasiparticle is persistent. We also examine the spectral densities in the paramagnetic phase and find that a spin polaron state can be excited, too. Its lifetime is smaller by several orders of magnitude than in the ferromagnetic phase but the ratio depends very sensitively on the values of $J$, $n$ and and the quasimomentum $k$. An interesting effect is found that the lifetime is maximal in the ferromagnetic phase if the quasimomentum is close to the Fermi volume.

	The paper is outlined as follows: In section \ref{sec::methods} we will discuss the method, particularly how we calculate spectral densities, how we extract the lifetimes and how we extrapolate them. In section \ref{sec::results} we will present our results. We will end up in a brief summary in section \ref{sec::discussion}.

	\section{Methods}
	\label{sec::methods}
	In this section we describe the methods used in our calculations. First of all we briefly discuss our DMRG algorithm. Secondly we describe the correction vector method, which we use to calculate Green's functions. At last we show how to calculate the lifetime of quasiparticles using the spectral functions we obtained from the Green's functions.
	
	\subsection{DMRG}
		For the calculation of ground states, we use a \emph{DMRG} algorithm with abelian and non-abelian symmetries, whose implementation is based on a matrix-product formulation. We use open boundary conditions for all calculations.
	
	\subsection{Correction vector method}
		Applying the \emph{correction vector}\cite{Soos1989,Ramasesha1997,Kuhner1999,Jeckelmann2002} method we obtain the spectral functions $A(\omega)$, where $\omega$ is the energy. To calculate $A(\omega)$, we need the retarded Green's function $G_{A}(\omega+i\eta)=G^+_{A}(\omega+i\eta)+G^-_{A}(\omega+i\eta) $, where 
		\begin{eqnarray}
		 G^+_{A}(\omega+i\eta) & =&  \bra{0} A^{\dagger} \frac{1}{\omega+ E_0 + i\eta - H} A \ket{0}\\
		 G^-_{A}(\omega+i\eta) & = & \bra{0} A \frac{1}{\omega - E_0 + i\eta + H} A^{\dagger} \ket{0}
		\end{eqnarray} 
		are the two branches of the retarded Green's function and $A$ is an arbitrary operator, $\ket{0}$ is the ground state with energy $E_0$  and $\eta>0$ is an artificial broadening factor, which is needed to lower the lifetime of the excitation to avoid boundary effects due to the finite system size. The basic rule is to choose $\eta > \frac{c}{L}$, where $c$ is the velocity of the excitation, but the minimal $\eta$ is strongly depending on the model.
		
		In principle one would need to compute both branches of the Green's function to obtain the complete spectral properties. For the determination of life times the spectral weight of the quasiparticle is nearly completely concentrated in one of the branches. Therefore we can neglect the other branch in this case.
		From now on, we will base all our arguments concerning the Green's function on the $+$-branch. Concerning the spectral density the calculations for the $-$-branch can be done similarly except for a minus sign.
		
		The correction vector is defined as
		\begin{eqnarray}
		 \ket{c(\omega+i\eta)} = \frac{1}{\omega+ E_0 + i\eta - H} A \ket{0}
		\end{eqnarray} 
		and hence
		\begin{eqnarray}
		 \left(\omega+ E_0 + i\eta - H \right)\ket{c(\omega+i \eta)} = A \ket{0},
		\end{eqnarray} 
		where the groundstate $\ket{0}$ ist obtained from the preceding DMRG calculation. This leads to a non-hermitian system of linear equations for real and imaginary parts, which can be solved using the GMRES method \cite{saad:856}. The outcome is $\ket{c(\omega + i\eta)}$, which allows to calculate the Green's function as
		\begin{eqnarray}
		 G_{A}(\omega+i\eta) = \bra{0} A \ket{c(\omega + i\eta)}.
		\end{eqnarray} 
		The spectral density can then be obtained by applying the standard formula
		\begin{eqnarray}
		 \label{eq::spectral-green}
		 A(\omega+i\eta) = - \frac{1}{\pi} \mbox{Im } G_{A}(\omega + i\eta),
		\end{eqnarray}
		where $\omega$ is assumed to be real.
		
	\subsection{Quasiparticle lifetime}
	\label{sec::quasiparticle-lifetime}
		For the calculation of \emph{quasiparticle lifetimes} we will limit ourselves to electronic systems. It is useful to transform the Hamiltonian into the fourier space according to
		\begin{eqnarray}
		 c_{i \sigma} = \frac{1}{\sqrt{N}}\sum_{k}c_{k \sigma} e^{i k r_i}.
		\end{eqnarray} 
		Hence we obtain
		\begin{eqnarray}
		 H = \sum_{k} \sum_{\sigma=\uparrow,\downarrow} \left( \epsilon_0(k) c_{k \sigma}^{\dagger} c_{k \sigma} \right) + J \sum_{k} \mathbf{S}_k \cdot \mathbf{s}_{-k}
		\end{eqnarray} 
		with $\epsilon_0(k)= -2t\cos{ka}$, where $a$ is the lattice spacing.
		The one electron Green's function is then defined as
		\begin{eqnarray}
		\label{eq::GF}
		 G_{k \sigma}(\omega+i\eta) = \bra{0} c_{k \sigma} \frac{1}{\omega+ E_0 + i\eta - H} c_{k \sigma}^{\dagger} \ket{0}.
		\end{eqnarray} 
		
		The self-energy $\Sigma_{\sigma}(k,\omega)$ is implicitly defined for the interacting system $H$ as
		\begin{eqnarray}
		 G_{k \sigma}(\omega+i\eta)=\frac{1}{\omega+i\eta-\left( \epsilon_0(k) - \mu + \Sigma_{\sigma}(k,\omega + i \eta) \right)} \notag,
		\end{eqnarray} 
		with $\mu$ the chemical potential. Note that $\eta$ appears also in the self-energy. This is nescessary, because $\lim_{\eta \rightarrow 0}$ will not be carried out in the numerical calculations. In general, the self-energy is a complex function $\Sigma_{\sigma}(k,\omega) = R_{\sigma}(k,\omega) + i I_{\sigma}(k,\omega)$. 
		Separation of real and imaginary part leads to
		\begin{widetext}
		\begin{eqnarray}
		\label{eq::greensfunction}
		 G_{k \sigma}(\omega+i\eta) = \frac{1}{\omega-\left( \epsilon_0(k) - \mu + R_{\sigma}(k,\omega + i \eta) \right) + i\left( \eta - I_{\sigma}(k,\omega + i\eta) \right)}.
		\end{eqnarray}
		\end{widetext}
		We now assume, that the self-energy is continuous and only weakly depending on $\omega$ in the vicinity of a resonance $\omega_{i \sigma}=\epsilon_0(k) - \mu + R_{\sigma}(k,\omega) |_{\omega=\omega_{i \sigma}}$, where $\omega_{i \sigma}$ is one out of several resonances, which are well separated to provide the correct determination of the lifetime of the quasiparticles (see the end of this section for the explicit extrapolation scheme). In addition we assume $|I_{\sigma}(k,\omega)| \ll | \epsilon_0(k) - \mu + R_{\sigma}(k,\omega) |$ near to the resonance we are interested in, i.e. we assume long lifetimes, because we are interested in these. This leads  to
		\begin{eqnarray}
			I_{\sigma}(k,\omega + i \eta) \approx I_{\sigma}^{(i)}(k),
		\end{eqnarray}
		in the vicinity of the $i$th resonance. For the real part of the self-energy we apply a Taylor expansion at the resonance $\omega_{i \sigma}$. We find
		\begin{eqnarray*}
			&&\omega - \left( \epsilon_0(k) - \mu + R_{\sigma}(k,\omega + i \eta) \right) \approx \\
			&\approx& \left( \omega - \omega_{i \sigma} \right) \left( 1- \frac{dR_{\sigma}(k,\omega+ i \eta)}{d \omega}|_{\omega+ i \eta=\omega_{i\sigma}} \right)\\
			& - &  i \eta \frac{dR_{\sigma}(k,\omega+ i \eta)}{d \omega}|_{\omega+ i \eta=\omega_{i\sigma}}
		\end{eqnarray*}
		and define
		\begin{eqnarray}
		 \alpha_{i \sigma}= \left( 1- \frac{dR_{\sigma}(k,\omega)}{d \omega}|_{\omega=\omega_{i \sigma}} \right)^{-1}.
		\end{eqnarray} 
		Substituting this to Eq. (\ref{eq::greensfunction}) the Green's function in the vicinity of resonance $\omega_{i \sigma}$ is given by
		\begin{eqnarray}
			G_{k \sigma}(\omega+i\eta) = \alpha_{i \sigma} \frac{1}{\omega- \omega_{i\sigma}  + i\left( \eta + \alpha_{i \sigma} \left|I_{\sigma}^{(i)}(k)\right| \right)}
		\end{eqnarray}
		and the spectral function obtains the form
		\begin{eqnarray}
		\label{eq::spectral_function}
		 A_{k \sigma}(\omega + i \eta)  &=& \notag\\ &&\hspace{-3cm}=\,\sum_i \frac{\alpha_{i \sigma}}{\pi} \frac{ \eta+ \alpha_{i \sigma} \left|I_{\sigma}^{(i)}(k)\right| } {(\omega-\omega_{i \sigma})^2+   \left( \eta+ \alpha_{i \sigma} \left|I_{\sigma}^{(i)}(k)\right| \right)^2} \,,
		\end{eqnarray} 
		which corresponds to a sum of Lorentz distributions at the resonances $\omega_{i \sigma}$ with a broadening of 
		\begin{eqnarray}
			\label{eq::broadening}
			B^{(i)}(\eta) = \eta+ \alpha_{i \sigma} \left|I_{\sigma}^{(i)}(k)\right| .
		\end{eqnarray} 
		Hence the broadening computed with the correction-vector method is essentially the sum of the natural broadening $\alpha_{i \sigma} \left|I_{\sigma}^{(i)}(k)\right|$ and the artificially introduced broadening $\eta$ and therefore $B(\eta)$ linearly depends on $\eta$.
		Note that from the Lehmann representation of the spectral density one can find that $I_{\sigma}^{(i)}(k) \leq 0$.
		
		The broadened spectral density $A_{k \sigma}(\omega + i \eta)$ is a convolution of the non-broadened spectral density $A_{k \sigma}(\omega)$ with a Lorentzian of width $\eta$:
		\begin{eqnarray}
		\label{eq::spectral_dens_lorentz}
		 A_{k \sigma}(\omega+i\eta) = \frac{1}{\pi}\int{d\omega' A_{k \sigma}(\omega) \cdot \frac{\eta}{(\omega - \omega')^2 + \eta^2}}.
		\end{eqnarray} 
		We now assume, that the spectral density consists of a sum of Lorentz distributions, which are separated by non-lorentzian regions. The outcome of the convolution of two Lorentzians again is a Lorentzian, where the broadenings behave additively.
		\begin{figure}[tb]
			\begin{center}
				\includegraphics[width=8.6cm]{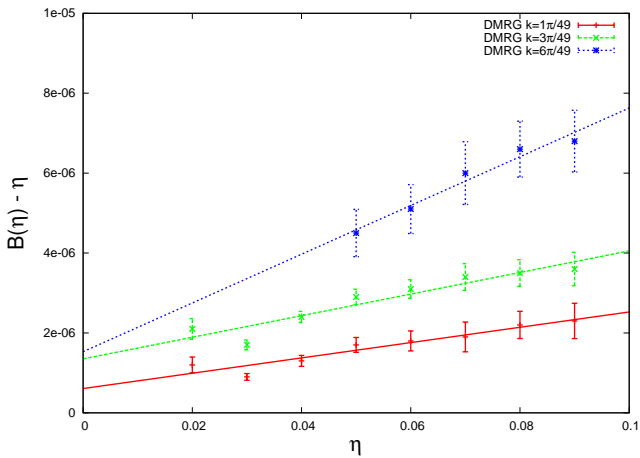}
				\caption{\emph{(Color online)} The linear fit of $B(\eta)$ vs $\eta$ is shown for three different quasimomenta, for a system with 48 sites, $n=0.125$ and Kondo coupling $J=1$. The data has been offset by $\eta$.}
				\label{fig::Broadening_vs_eta_48Sites_18e_J05}
			\end{center}
		\end{figure}
		As the broadening corresponds to an inverse lifetime, we can define the lifetime of the  quasiparticle corresponding to the $i$th resonance as
		\begin{eqnarray}
		\label{eq::lifetime}
		 \tau &= & \lim_{\eta \rightarrow 0} \frac{1}{ \eta + \alpha_{i \sigma} \left| I_{\sigma}^{(i)}(k) \right| } \notag\\
		 & = &\frac{1}{\alpha_{i \sigma} \left|I_{\sigma}^{(i)}(k)\right|}.
		\end{eqnarray} 
		The limitations of this method are obvious. First of all the excitation must cause a lorentzian shaped peak in the spectral density. To be able to extract the broadening of such a peak all other peaks must be energetically separated from this one. Thus we have to check, whether the conditions of our theory are fulfilled or not. We can check whether the spectral density has a lorentzian shape (see Fig. \ref{fig::perfect_lorentz-fit}) and $B(\eta)$ has to depend linearly on $\eta$ (see Fig. \ref{fig::Broadening_vs_eta_48Sites_18e_J05}).
		
		We have now two possible estimates for the lifetime of a quasiparticle:
		\begin{enumerate}
		 \item Lower estimate: Use the inverse broadening $B(\eta)^{-1}$ directly (without $\eta \rightarrow 0$). With (\ref{eq::lifetime}) $B(\eta)^{-1}$ is smaller than $B(0)^{-1}$, this is therefore a reliable lower estimate.
		 \item Extrapolation: Calculate the broadening for several different $\eta$. From this one can extract the linear dependence of $B(\eta)$ on $\eta$ and $B(0)^{-1}$ gives the extrapolated lifetime. See also Fig. \ref{fig::Broadening_vs_eta_48Sites_18e_J05}.
		\end{enumerate}
		\begin{figure}[tb]
			\begin{center}
				\includegraphics[width=8.6cm]{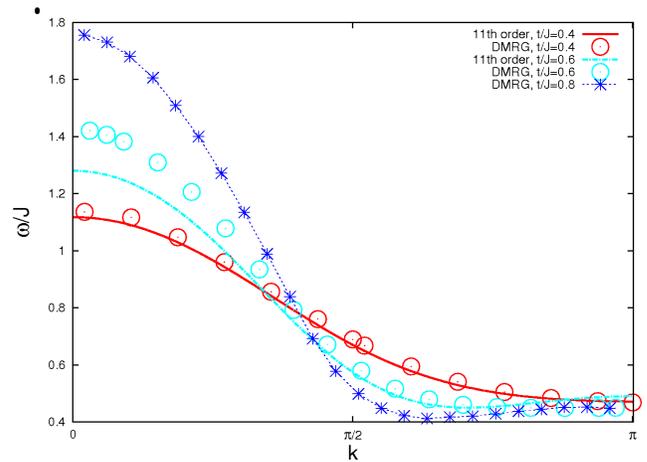}
				\caption{\emph{(Color online)} Dispersion relations of the half-filled KLM. The comparison to 11th order perturbation theory is taken from [\onlinecite{Trebst2006}]. The line for $t/J=0.8$ is meant only as a guide to the eye.}
				\label{fig::dispersion_half_filling}
			\end{center}
		\end{figure}
		It turns out that due to long lifetimes only extrapolated lifetimes are meaningful. Therefore in the next section we will discuss the results obtained by the second method only.
		
		From the Lorentzian fit of a single resonance peak of the spectral function in Eq. (\ref{eq::spectral_function}) one can also estimate the spectral weight $\alpha_{i \sigma}$ of the corresponding excitation.

	\section{Results}
	\label{sec::results}
		\begin{figure}[tbh]
			\begin{center}
				\includegraphics[width=8.6cm]{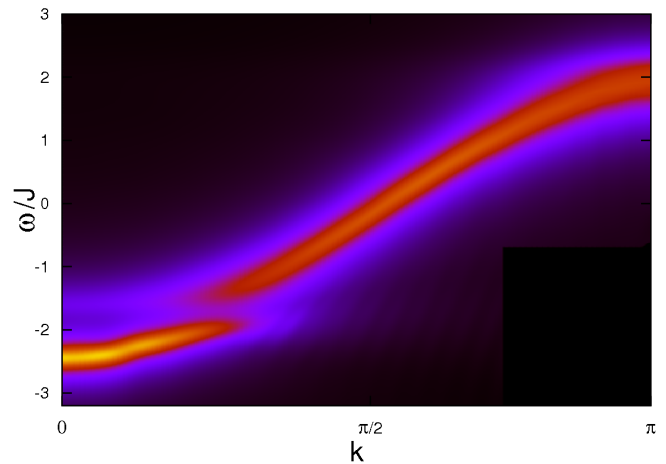}
				\caption{\emph{(Color online)} Dispersion of a KLM with 48 Sites, $n=0.125$ and $J=1$. The broadening is $\eta=0.2$. The lower band is the bound quasiparticles band, while the upper one is the scattering states band.}
				\label{fig::dispersion_48Sites_6e_map}
			\end{center}
		\end{figure}
		In this section we will present the results obtained using the methods we discussed in the last section. First we will show the calculated dispersion relations considering a Kondo lattice model at half-filling and at partial filling. Our half-filling results show a qualitative agreement with the results in Ref. \onlinecite{Trebst2006}. In the second part we show several spectral functions and the calculated lifetimes, which leads to the conclusion that we find a bound polaron state. The hopping parameter $t$ is set to $t=1$ in all calculations.
		
	\subsection{Dispersion}			
		The half filled KLM serves as the touchstone of our method, where we can compare our results to those of Trebst et al. \cite{Trebst2006}, who did a strong coupling expansion up to 11th order in $t/J$. We calculated the dispersion relation for different values of $t/J$, see Fig. \ref{fig::dispersion_half_filling} and used lattice sizes of 32 and 48 sites. The calculations have converged in the sense that we could not find any deviations between calculations of different system sizes. Our results show very well agreement to the results in Ref. \onlinecite{Trebst2006} for $t/J=0.4$ with small deviations for small $k$. By strong coupling expansion it is found that the band flattens out for $k \rightarrow \pi$ around $t/J_c \approx 0.50 \pm 0.02$ and therefore the effective quasiparticle mass diverges. This is also found by DMRG for a higher value of $t/J_c \approx 0.576 \pm 0.002$. As one can see, in Fig. \ref{fig::dispersion_half_filling}, the strong coupling expansion dispersion relation has a minimum at $k < \pi$ for $t/J=0.6$, which is not found by DMRG. DMRG exhibits this minimum again only for higher values of $t/J$. In Fig. \ref{fig::dispersion_half_filling} it is for instance shown for $t/J=0.8$. Summarizing, the DMRG results match very well to the strong coupling expansion for low $t/J$ but the agreement becomes worse for $t/J \geq 0.6$. DMRG is the more reliable method in that regime, because it is non-perturbative and the error can be easily controlled by very small DMRG truncation errors. In this case it is easy to keep the truncation error reasonably low. We can confirm the physical picture established by Trebst et al., namely that the quasiparticles gain an enormously high mass, which is due to a growing correlation between the conduction and the f-electrons. The quasiparticles with high momenta therefore behave like almost localized f-electrons.

		Now we consider the dispersion relation of the KLM for partial band filling, see Fig. \ref{fig::dispersion_48Sites_6e_map}. For the ground state calculation of the KLM with 48 sites, a filling of $n=0.125$ and $J=1$ we used about 100 DMRG ansatz states. The calculation of the correction vectors needed 800 DMRG states to reach good convergence. In Fig. \ref{fig::dispersion_48Sites_6e_map} and all other figures of dispersion relations we neglected the chemical potential, which would shift the lower band edge to nearly zero.
		\begin{figure}[btp]
			\begin{center}
				\includegraphics[width=8.6cm]{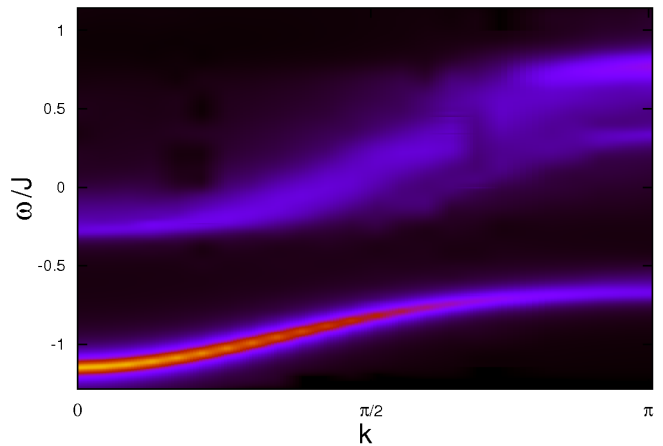}
				\caption{\emph{(Color online)} Dispersion relation of a KLM with $n=0.125$ and $J=3.5$. The lower band is the bound quasiparticles band, while the upper one is the scattering states band.}
				\label{fig::dispersion_n0125_J2u3}
			\end{center}
		\end{figure}
		One can distinguish two different bands. The higher one behaves like $2t \cos{k}+2t$ and can therefore be attributed to free electrons, which do not form bound states with the localized spins. From now on, this band will be referred to as the \emph{scattering states band} in the sense that these excited states rapidly decay. The lower one of the two bands represents the states of the system which are formed by the conduction electrons bound to the localized spins, from now on referred to as \emph{quasiparticle or Spin-polaron band}. Contrary to the scattering states band this band consists only of one state, which is separated from the continuum (for large $N$) of scattering states and has a Lorentzian shape from which the lifetime can be extracted, which is very long in most of the cases, see Sec. \ref{sec::life_times}. Keeping the same filling $n=0.125$, but raising the Kondo coupling constant $J$ to $3.5$, see Fig. \ref{fig::dispersion_n0125_J2u3}, the quasiparticle band becomes more separated from the scattering states, because the quasiparticle state is now energetically lowered. 
		This can be understood by a simple physical picture. For that we rewrite the exchange coupling of the Hamiltonian as $\sum_i^L \left[ J^z S_i^z s_i^z + \frac{J_{\perp}}{2} \left( S_i^+ s_i^- + S_i^- s_i^+ \right) \right]$ and we now set $J_{\perp}=0$. The quasiparticle state of the KLM almost only consist of an electron with spin antiparallel to the localized spins. With respect to our modified exchange coupling this results in a lowered energy of $J^z/4$ per electron. The scattering states also contain important contributions with an electron spin oriented antiparallel to the localized spins. This leads to a higher energy of $J^z/4$.
		Therefore the energy difference between quasiparticle and scattering states is $J^z/2$ and scales with $J^z$. Taking also a finite $J_{\perp}$ into account the quasiparticle energy is even lowered more due to spin flip processes. 
		The scattering states band has a similar shape as before, as expected. The weight of the quasiparticle states band is also increasing with $J$. This is also expected, because the state becomes energetically more favorable with increasing $J$.
		
		\begin{figure}[tbp]
			\begin{center}
				\includegraphics[width=8.6cm]{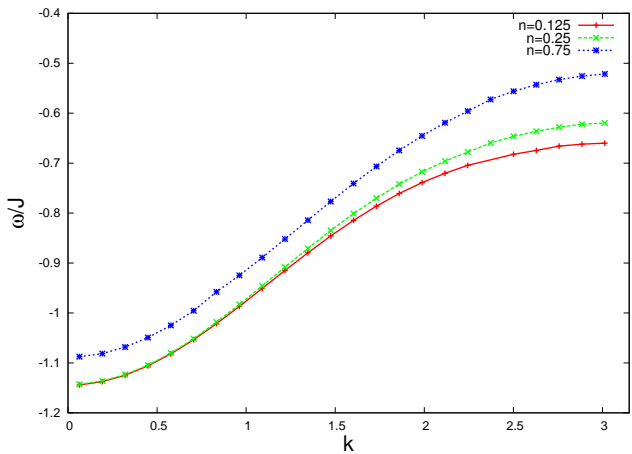}
				\caption{\emph{(Color online)} Dispersion relations of a KLM with $J=3.5$ and three different fillings, $n=0.125$, $n=0.25$ and $n=0.75$. The \emph{inset} shows the second derivative of the dispersion relation, which is $\propto \frac{1}{m^*}$, where $m^*$ is the effective mass.}
				\label{fig::dispersion_J35}
			\end{center}
		\end{figure}
		In Fig. \ref{fig::dispersion_J35} we show the dispersion relation of the quasiparticle of a system with $J=3.5$ for three different fillings, $n=0.125$, $n=0.25$ and $n=0.75$. The ground state is ferromagnetic in all cases. We conclude that even in the presence of many electrons the spinpolaron state can be clearly identified.
		
		\begin{figure}[bp]
			\begin{center}
				\includegraphics[width=8.6cm]{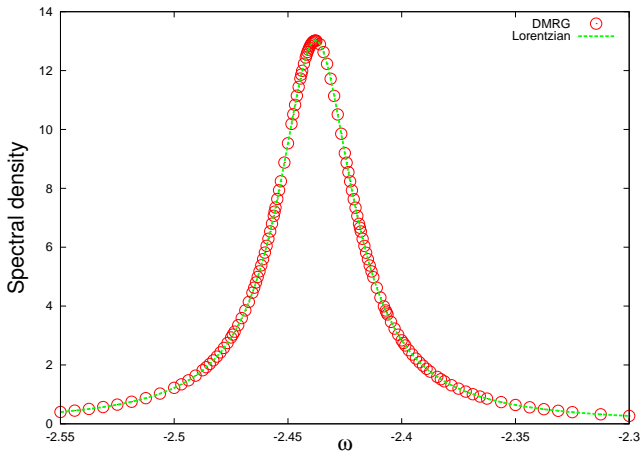}
				\caption{\emph{(Color online)} Spectral function of a KLM with 48 Sites, $n=0.125$, $J=1$, $k=\frac{\pi}{49}$ and $\eta=0.02$. $L_{\eta \omega_0}(\omega)$ is a Lorentzian fitted to the DMRG values.}
				\label{fig::perfect_lorentz-fit}
			\end{center}
		\end{figure}

		\subsection{Life time estimations from spectral functions}
		\label{sec::life_times}
		In a further step we take a look at single spectral densities for fixed quasimomentum $k$, which provides the possibility to calculate quasiparticle lifetimes of the bound quasiparticles and prove the existence of bound polaron states. We consider only the calculation of the extrapolated lifetimes, as described in section \ref{sec::quasiparticle-lifetime} and whose extrapolation scheme is shown in Fig. \ref{fig::Broadening_vs_eta_48Sites_18e_J05}. Calculating extrapolated lifetimes this way we have to be very careful due to the assumption that the spectral density complies a lorentzian distribution. For the spectral density being a lorentzian the imaginary part of the self-energy has to be very small compared to the energy of the resonance and it should not vary too much in the vicinity of the resonance. The expansion of the self energy leads then to a lorentzian function. Thus the spectral density is not lorentzian shaped in higher orders of the expansion and it has to be checked, see for example Fig. \ref{fig::perfect_lorentz-fit}, whether it is good enough. Fig. \ref{fig::perfect_lorentz-fit} shows a spectral density for a KLM with 48 Sites, $n=0.125$, $J=1$ and  quasimomentum $k=\frac{\pi}{49}$. The artificial broadening is set to $\eta=0.02$. The number of data points obtained provides the possibility of a very precise fit of the lorentz distribution. Fig. \ref{fig::perfect_lorentz-fit} sharply supports the assumption made in sec. \ref{sec::quasiparticle-lifetime} that the spectral density has a lorentzian shape, which is nescessary to calculate quasiparticle lifetimes.

		\begin{figure}[tbp]
			\begin{center}
				\includegraphics[width=8.6cm]{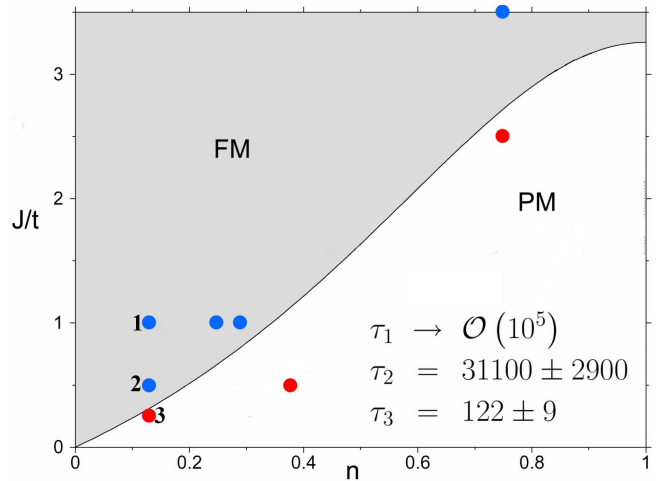}
				\caption{\emph{(Color online)} Simplified phase diagram of the 1D Kondo lattice model taken from [\onlinecite{McCulloch2001}]. The points mark the parameters at which extrapolated lifetimes have been calculated. The lifetimes for points 1,2 and 3 and quasimomentum $k=\frac{\pi}{49}$ are given directly in the picture by $\tau_1$, $\tau_2$ and $\tau_3$, the lifetimes for the other points are listed in tab. \ref{tab::extrapolated_life_times}}
				\label{fig::phase_diagram}
			\end{center}
		\end{figure}		

		The extrapolated lifetimes are summarized in Fig. \ref{fig::phase_diagram} and Tab. \ref{tab::extrapolated_life_times}. There we can see, that the lifetime strongly depends on the parameters filling $n$ and Kondo coupling constant $J$ as well as on the quasimomentum $k$.
		The lifetimes in the ferromagnetic phase (this concerns the $\left[n,J\right]$-pairs $\left\{ [0.125,0.5], [0.125,1], [0.25,1], [0.29,1], [0.75,3.5]\right\} $ ) decrease by approaching the paramagnetic phase by either lowering $J$ or increasing $n$. For fixed and low quasimomentum $k$ it seems that the life-time decreases by increasing $n$ (even if $J$ is increased at the same time so that the distance to the paramagnetic phase is still large, compare e.g. the pairs $[0.125,1]$ and $[0.75,3.5]$). This indicates that the life-time is influenced by the presence of other quasiparticles, probably by an effective interaction between the quasiparticles mediated via the coupling to the localized spins. This is further substantiated by the dependence of the quasiparticle life-time on the quasimomentum $k$ in the ferromagnetic phase. For $k$ approaching the Fermi level, the lifetime increases, which is consistent with the fact that the phase space for quasiparticle interaction becomes smaller close to the Fermi level. In contrast, electron-magnon interaction is expected to be more important for quasiparticles close to the Fermi-level because the energy of the spinpolaron comes closer to the scattering band. This effect can be seen in the paramagnetic phase for the pairs $[0.375,0.5]$ and $[0.75,2.5]$, where the lifetime decreases with increasing quasimomentum. Thus, in the paramagnetic phase, we conclude that electron-magnon interaction limits the life-time of the spinpolaron. Deep inside the paramagnetic phase at $[n=0.375,J=0.5]$ the lifetime is short for all determined values of $k$. Therefore, as predicted earlier in Refs. [\onlinecite{PhysRevLett.78.2180,PhysRevB.65.052410}] there exist no persistent quasiparticles in this regime.
		
		We also extracted the spectral weight of the spin polaron excitation from the Lorentzian fit and summarized them in Tab. \ref{tab::extrapolated_life_times} in the second row of the respective $k$ value. Considering the three numbered points of Fig. \ref{fig::phase_diagram} we calculated the spectral weights
		\begin{enumerate}
			\item $0.818004 \pm 0.000001$
			\item $0.87119 \pm 0.00003$
			\item $0.588 \pm 0.001$,
		\end{enumerate}
		which do fulfill the expectation that the spectral weight should be significantly lower in the paramagnetic phase. The calculated weights are independent of $\eta$ within the errorbounds. They show a strong dependence on the quasimomentum (decreasing for growing $k$) in the ferromagnetic as well as in the paramagnetic phase. This is expected because the spinpolaron states with higher value of $k$ have higher energy and come closer to the scattering states. However, it is unexpected that the spectral weight is large for $[0.375, 0.5]$ and this still has to be explained.
		
		\begingroup
		\squeezetable
		\begin{table}[tb]
		\begin{ruledtabular}
		\caption{Summarization of extrapolated lifetimes (first row of the respective k value) and spectral weight of the spin polaron excitation (second row of the respective k value) for different fillings of the system, different coupling constants and different quasimomenta. The $k$ values in paranthesis correspond to $[0.75,2.5]$.}
			\begin{tabular}{l p{1.3cm} p{1.1cm} p{1.5cm} p{1.3cm} p{1.3cm}}
				\backslashbox{k}{$[n,J]$}	& $[0.25, 1]$ & $[0.29, 1]$ & $[0.375, 0.5]$ & $[0.75, 3.5]$ & $[0.75, 2.5]$\\
			\hline 
				$\frac{1\pi}{49}$ $\left( \frac{1\pi}{33} \right)$	& $831 \pm 16$ & $183 \pm 3 $ & $67 \pm 9$ & $38.6 \pm 0.8$ & $25.7 \pm 1.6$\\	
					& $0.8211 \pm 0.0004$ & $0.8292 \pm 0.0008$ & $0.951 \pm 0.006$ & $0.714 \pm 0.008$ & $0.674 \pm 0.005$\vspace{0.3cm}\\
				$\frac{3\pi}{49}$ & $2540 \pm 84$ & & &\vspace{0.3cm}\\
				$\frac{8\pi}{49}$	& & & $14.7 \pm 1.5$ & &\\
					& & & $0.82 \pm 0.01$ & &\vspace{0.3cm}\\
				$\frac{11\pi}{49}$ & $\mathcal{O}\left( 10^5 \right) $ & $2850 \pm 600$ &  & &\\
					& $0.6244 \pm 0.0001$ & $0.6375 \pm 0.0005$ &  & &\vspace{0.3cm}\\
				$\frac{18\pi}{49}$ $\left( \frac{12\pi}{33} \right)$  & & & & $\mathcal{O}\left(10^3 \right)$& $5.29 \pm 0.14$\\
					& & & & $0.5580 \pm 0.0008$& $0.3391 \pm 0.0005$\vspace{0.3cm}\\
				$\frac{36\pi}{49}$ $\left( \frac{24\pi}{33} \right)$  & & & & $\mathcal{O}\left(10^3 \right)$ & $12.76 \pm 0.62$\\
					& & & & $0.3391 \pm 0.0005$ & $0.190 \pm 0.006$\\
			\end{tabular}
			\label{tab::extrapolated_life_times}
			\end{ruledtabular}
		\end{table}
		\endgroup

	\section{Summary}
	\label{sec::discussion}
	We have studied the one-dimensional Kondo lattice model at half-filling and at partial band fillings for various Kondo couplings $J$. At half-filling we could verify the results of Ref. \onlinecite{Trebst2006}. This includes the dispersion relation and the divergence in the effective quasiparticle mass. At partial band fillings we were able to show that in the case of ferromagnetism long living quasiparticle states exist, the spin polaron quasiparticle. The lifetime exceeds the lifetime of quasiparticle excitations deep inside the paramagnetic phase by several orders of magnitude. From the dependence on the quasimomentum we conclude that the dominant decay process is the spinpolaron-spinpolaron interaction in the ferromagnetic phase, and the interaction between electrons and spin waves in the paramagnetic phase. The weight of the spinpolaron state is very close to one even for special points in the paramagnetic phase. The results motivate the speculation that spin coherence can be significantly enhanced by coupling of electrons to magnons in the ferromagnetic phase of the localized spins. Even if the localized spins have a negligible exchange interaction, the itinerant electrons can mediate an effective exchange interaction which drives the local spins into a ferromagnetic phase at low enough temperatures (the critical temperature has been estimated recently and is within experimental reach.\cite{Simon2008, Braunecker2008,Braunecker2008a}) As we have seen in this paper, the spin excitations in the ferromagnetic phase can in turn form spinpolaron bound states with the itinerant electrons, increasing their life-time considerably. This effect persists in the presence of many electrons and becomes more efficient for quasimomenta close to the Fermi level. It is an interesting question for future research to investigate the consequences for the spin relaxation and dephasing rates in the Kondo lattice model by studying the spin-spin correlation functions.

	\begin{acknowledgements}
          S. Smerat, Ulrich Schollwöck and H. Schoeller acknowledge the support from the 
          DFG-Forschergruppe 912 on \lq\lq Coherence and relaxation properties of electron spins''.
	\end{acknowledgements}

	\bibliographystyle{prb}

	
\end{document}